\documentclass[aps,prd,twocolumn,groupedaddress]{revtex4}

\usepackage{amsmath}
\usepackage{graphicx}
\usepackage{subfigure}
\usepackage{epstopdf}

\newcommand{\beq}{\begin{equation}}
\newcommand{\eeq}{\end{equation}}
\newcommand{\ov}{\overline}

\usepackage{color}

\begin{document}


\title{Light dark matter in leptophobic $Z'$ models}



\author{P. Gondolo}
\email[]{paolo.gondolo@utah.edu}
\affiliation{Department of Physics, University of Utah, Salt Lake City, UT }
\altaffiliation{School of Physics, KIAS, Seoul 130-722, Korea}

\author{P. Ko}
\email[]{pko@kias.re.kr}
\affiliation{School of Physics, KIAS, Seoul 130-722, Korea}

\author{Y. Omura}
\email[]{omura@kias.re.kr}
\affiliation{School of Physics, KIAS, Seoul 130-722, Korea}


\date{\today}

\begin{abstract}
\noindent
Recent experimental results in direct dark matter detection may be interpreted in terms of a dark matter particle of mass around 10 GeV$/c^2$. We show that the required scenario can be realized with a new dark matter particle charged under an extra abelian gauge boson $Z'$ that couples to quarks but not leptons. This is possible provided the $Z'$ gauge boson is very light, around 10--20 GeV$/c^2$ in mass, and the gauge coupling constant is small, $\alpha'\sim10^{-5}$. Such scenarios are not constrained by accelerator data. 
\end{abstract}


\maketitle

\section{Introduction}

Many astrophysical observations support the existence of cold dark matter (CDM) in our universe,
but its properties, such as its mass, spin, and interactions, are still largely unknown. Several decades of experimental work have been spent trying to detect dark matter particles coming to Earth (direct detection), or more generally products of dark matter reactions in outer space (indirect detection). In these dark matter searches, one must differentiate signals due to dark matter from signals of other origin. In direct searches, a paramount telltale sign is a regular variation of the particle counting rate with a period of one sidereal year: an annual modulation \cite{Drukier:1986tm}. Since 1998, the DAMA experiment has been detecting such a modulation \cite{Bernabei:1998fta}, with a current significance of more than 8$\sigma$~\cite{Bernabei:2010mq}. Other direct detection experiments have placed strong constraints on the interpretation of the DAMA modulation in terms of dark matter particles (see e.g.~\cite{Savage:2010tg} for a recent analysis). 

A possibility that is not clearly excluded is a weakly-interacting massive particle of mass around a few GeV$/c^2$, as proposed by Gondolo and Gelmini \cite{Gelmini:2004gm}. This is a natural mass range to explain the cosmic density of cold dark matter by means of thermal production in the early-universe plasma \cite{Lee:1977ua}. Recently, this possibility has become quite enticing thanks to reports of excess counts of unknown origin in the CoGeNT~\cite{Aalseth:2010vx} and CRESST \cite{Probst:2010} experiments, and the hint that the CoGeNT events are modulated in a way similar to DAMA's \cite{Collar:2011}. Although these results are in tension with those of CDMS \cite{Ahmed:2010wy}, XENON10 \cite{Angle:2011}, and XENON100 \cite{Aprile:2011}, calibration uncertainties in the energy scale allow for a common interpretation of all the current measurements in terms of a $\sim$7 GeV$/c^2$ dark matter particle with a $\sim$10$^{-40}$ cm$^2$ scattering cross section with nucleons \cite{Hooper:2010uy}.

Particle physics models that include a light dark matter particle with the properties just discussed must be extensions of the Standard Model (SM). The minimal supersymmetric extension (MSSM), very light neutralinos might be possible at the edge of the available parameter space (see \cite{Bottino}, but see also \cite{Zurek} for  opposite opinions). Other extensions of the SM may offer less tight possibilities (see e.g.~\cite{Gunion}). In these and other models \cite{Kim:2009ke}, the interaction of the dark matter particle with quarks is (dominantly) mediated by the exchange of a  light Higgs boson. Models with extra abelian gauge bosons $Z'$ offer other possibilities for GeV dark matter. In these models, the coupling between the dark matter and ordinary matter could be achieved through kinetic mixing of the $Z'$ with the SM photon and Z boson \cite{Kang:2010mh}, through exchange of extra fermions \cite{Feng:2008dz}, or through the exchange of the $Z'$ boson itself \cite{Buckley:2010ve,Ko:2010at}. 

In this paper, we examine $Z'$ models in which the coupling of dark and ordinary matter is achieved through the exchange of the $Z'$ boson itself. It turns out that in these models the DAMA/CoGeNT region is compatible with a correct dark matter density when the dark matter annihilates in the early universe through a $Z'$ resonance. This requires (\cite{Gondolo:1990dk,Griest:1990kh}) a relic density calculation more careful than previous ones. We find that the appropriate $Z'$ boson should be light, $\sim$10--20 GeV$/c^2$, compatible with the supersymmetric U(1)$_B$ model in \cite{Ko:2010at} but in contrast to the work of \cite{Buckley:2010ve} who require a $\sim$150 GeV$/c^2$ $Z'$ boson to describe the CDF dijet anomaly.

\section{Extra U(1)$'$ gauge boson as a mediator}
\label{sec;gauge}

We consider an extra U(1)$'$ gauge group under which  the CDM particle and  some standard model particles are charged. The U(1)$'$ symmetry is spontaneously broken according to a nonzero vacuum expectation value (vev) of a Higgs field $\phi$.
Then the massive U(1)$'$ gauge boson becomes a mediator between 
the CDM sector and the SM sector.  

The couplings of the $Z'$ boson to SM particles must be chosen wisely. If the SM leptons are charged under U(1)$'$ 
in a universal way, there is a strong constraint from LEP-II and Tevatron~\cite{Carena:2004xs} on the ratio  $m_{Z'}/g'$ between the mass $m_{Z'}$ of the extra gauge boson and the coupling constant $g'$ of the extra {\rm U(1)} gauge group. For example, for U(1)=U(1)$_{B-L}$, the LEP-II bound is approximately $m_{Z'}/g' \gtrsim 6$ TeV$/c^2$. The constraint can be relaxed if electrons are neutral under U(1)$'$. 
If only muons, taus and the CDM particles are charged under U(1)$'$, one can construct
a viable leptophillic dark matter model that accounts for some excess events observed in indirect dark matter searches \cite{baek_ko}, but the required dark matter mass is much heavier than the few GeV$/c^2$ we seek here. If only quarks and the CDM are charged (leptophobic $Z'$), the LEP-II/Tevatron constraints are even more relaxed \cite{Carone:1995,Carone:1995pu,Aranda:1998fr}, and as we show in this paper, it is possible to have a $Z'$ model with a light dark matter particle. Finally, the $Z'$ boson can couple to quarks and leptons indirectly through kinetic mixing with the SM photon and $Z$ boson. Although such kinetic mixing is strongly constrained by experiments \cite{Carone:1995pu,Aranda:1998fr}, it can lead to viable models for dark matter interactions \cite{Chun:2010ve,Kang:2010mh}.

Here we present two $Z'$ models with a viable light CDM candidate: in one, the CDM particle is a scalar boson, in the other, it is a Dirac fermion.
We realize the coupling to ordinary matter needed for direct CDM detection through the exchange of a leptophobic $Z'$ (couplings of the $Z'$ with the $\tau$ lepton are in principle possible too). Although we impose no $Z'$-$\gamma$ or $Z'$-$Z$ kinetic mixing, a small amount of kinetic mixing could be generated radiatively. However, it turns out that in our scenario the gauge coupling constant $g'$ is small ($\sim$10$^{-2}$) and any radiatively-generated kinetic mixing is highly suppressed.


\subsection{Scalar CDM with $Z'$ mediator}
\label{sec;scalar}

Here we  consider a scalar CDM particle $X$ charged under the new U(1)$'$ gauge group.
We add the following lagrangian to the standard model,
\begin{align}
{\cal L}'_{\rm scalar} & = 
D_\mu X^\dagger \,D^\mu\! X - m_X^2 X^\dagger X 
- \frac{\lambda_X}{4} (X^\dagger X)^2
\nonumber \\ & 
+ D_\mu \phi^\dagger D^\mu \phi - m_\phi^2 \phi^\dagger \phi 
- \frac{\lambda_\phi}{4} (\phi^\dagger \phi)^2
\nonumber \\ &
- \frac{\lambda_{HX}}{2} X^\dagger X H^\dagger H 
- \frac{\lambda_{X\phi}}{2} \phi^\dagger \phi X^\dagger X
\nonumber \\ &
- \frac{\lambda_{H\phi}}{2} \phi^\dagger \phi H^\dagger H  
-\frac{1}{4}  Z'_{\mu\nu} Z'^{\mu\nu}.  
\end{align}
We also add a U(1)$'$ term to all standard-model covariant derivatives $D_\mu^{\rm SM}$,
\beq
D_\mu = D_\mu^{\rm SM} - i Q' g' Z'_\mu.
\eeq
Here $g'$ is the U(1)$'$ gauge coupling constant and $Q'$ is
the U(1)$'$ charge of the field on which $D_\mu$ acts. We will also use $\alpha'=g'^2/(4\pi)$. We assume that the U(1)$'$ gauge boson couples to the SM 
fermions vectorially, and that the gauge anomaly is canceled by new fields charged under U(1)$'$ \cite{Ko:2010at,wise}.

In the above lagrangian, $H$ is the SM Higgs boson, and $\phi$ is the U(1)$'$ Higgs boson whose vev breaks U(1)$'$.
In order to realize $\langle \phi \rangle \neq 0$ and $\langle H \rangle \neq 0$, at least $m_{\phi}^2$ and $m_{H}^2$ must be negative. The $\lambda_{H \phi}$ term stabilizes the vacuum, and
the negative squared masses must be tuned in correspondence with the size of the $\lambda_{H \phi}$ term.

This type of a model has been studied by Wise et al.~\cite{wise}.
In their model, U(1)$'$ is a gauged baryon symmetry ${\rm U(1)}_B$, and $Q'_X$ is fixed by the Yukawa coupling that allows non-SM charged particles to decay.
We will later discuss the ${\rm U(1)}_B$ case as a concrete example.

For the stability of the CDM particle $X$ on cosmological timescales, the field $X$ should not acquire a nonzero vev $\langle X \rangle$, because a nonzero $\langle X \rangle$ induces the trilinear coupling  $\lambda_{HX}\langle X \rangle X H^\dagger H$, which allows for decay processes such as $X \rightarrow H^{\dagger}H$. Further terms that cause the $X$ particles to decay, such as $X \phi^{-Q'_X/Q'_\phi}$, arise from one-loop and non-renormalizable corrections. One can impose that the stability of the $X$ particle at the renormalizable level be means of the conditions
\beq
Q'_X \neq \pm 2 Q'_\phi,~ 3 Q'_\phi,
\eeq
but one cannot completely forbid $X$ to decay through higher-order couplings. 
 We assume that $m_X^2 >0$ is satisfied and that higher-order unsafe couplings are small enough to guarantee a life time for the $X$ particles comparable or greater than the age of the universe.
In our numerical calculations, we take $Q'_X=1$, $Q'_q=1/3$ for quarks, and $Q'_l=0$ for leptons, which is typical of a U(1)$_B$ coupled to baryon number.

The annihilation cross section for $XX^*\to f\ov{f}$ through $Z'$ exchange is
\begin{align}
\sigma_{X\ov{X}\to f\ov{f}} = 
\frac{8\pi Q'^2_X Q'^2_f  \alpha'^2 \beta\beta'(2E^2+m_f^2)}{(4E^2-m_{Z'}^2)^2+m_{Z'}^2\Gamma_{Z'}^2}.
\end{align}
Here  $Q'_f$ is the U(1)$'$ charge of SM particle, $f$, with mass $m_f$,
$E$ is the $X$ or $f$ center-of-mass energy, and $\beta=\sqrt{1-m_X^2/E^2}$ and $\beta'=\sqrt{1-m_f^2/E^2}$ are the center-of-mass velocities of each $X$ and $f$ in the center of mass frame. In the low-velocity limit, the leading term is a p-wave.
The decay width of the $Z'$ boson is given by 
\begin{align}
\label{eq:width}
\Gamma_{Z'} &= \frac{\alpha'}{m_{Z'} }   \sum_{f} Q'^2_f \left( m_{Z'}^2 + 2m_f^2 \right) \sqrt{1 - \frac{4m_f^2}{m_{Z'}^2} }
\nonumber\\&+\frac{ \alpha'}{12m_{Z'} } Q'^2_X \left( m_{Z'}^2 -4 m_X^2 \right)  \sqrt{1 - \frac{4m_X^2}{m_{Z'}^2} }.
\end{align}
In principle, the CDM particle $X$ can annihilate to quarks not only through $Z'$ exchange, but also through the Higgs bosons $H$ and $\phi$. For simplicity we neglect the Higgs boson contributions under the assumption that either the Higgs bosons are heavy or their couplings $\lambda_{X\phi}$ and $\lambda_{HX}$ are small.
 
In addition, for $m_X>m_{Z'}$, the annihilation into $Z'$ pairs is allowed, with cross section
\begin{align}
&\sigma_{X\ov{X}\to Z'Z'} = \frac{\pi Q'^4_X\alpha'^2}{2E^2} \frac{w}{v}
\left[ \frac{32-24z^2+5z^4+16v^2}{4-4z^2+z^4+4v^2}
\right. \nonumber \\ & \left.
-\frac{16-8z^2-z^4+16v^2(2-z^2)}{4vw(1+v^2+w^2)} 
\ln\frac{1+(v+w)^2}{1+(v-w)^2}
\right].
\end{align}
Here $z=m_{Z'}/m_X$, $v=p/m_X$, and $w=k/m_X$, where $E$ and $p$ are the center-of-mass energy and momentum of the initial $X$ particles and $k=\sqrt{E^2-m_{Z'}^2}$ is the momentum of the final $Z'$ particles.

\subsection{Dirac fermion CDM with $Z'$ mediator}
\label{sec;Dirac}

Here we discuss the case of light CDM being a Dirac fermion charged under U(1)$'$. 
The SM is augmented by the following lagrangian terms.
\begin{align}
{\cal L}'_{\rm fermion} & =
\ov{\psi}_X \left( i \,/\llap{$\partial$} + g' Q'_X \rlap{\,/}Z' - m_X \right) \psi_X 
\nonumber  \\ &
+ D_\mu \phi^\dagger D^\mu \phi 
- m_\phi^2 \phi^\dagger \phi 
-  \frac{\lambda_\phi}{4} (\phi^\dagger \phi)^2
\nonumber  \\ &
- \frac{\lambda_{H\phi}}{2} \phi^\dagger \phi H^\dagger H
-  \frac{1}{4}  Z_{\mu\nu}' Z^{'\mu\nu} .
\label{eq:model}
\end{align}
Here, $\psi_X$ is a Dirac fermion with U(1)$'$ charge $Q'_X$.  
A global symmetry $\psi_X \rightarrow e^{i\theta } \psi_X$ can be enforced after the U(1)$'$ symmetry breaking,
so that $\psi_X$ is guaranteed to be stable. However, higher-order non-renormalizable terms could generally break the global symmetry. Such terms are for example $udd \ov{\psi}_X/\Lambda^2$ if U(1)$'={\rm U(1)}_B$ and $Q'_X=1$, and $N_R^3\psi_X/\Lambda^2$ if a right-handed neutrino $N_R$ is added and charged under U(1)$'$ with $Q'_X=1$ and $Q'_{N_R}=-1/3$. We assume that the cut-off scale $\Lambda$ is large enough that the $X$ particles are cosmologically stable.

The annihilation cross section for $X\ov{X}\to f\ov{f}$ through $Z'$ exchange is
\begin{align}
\label{eq;Diracani-sigma}
\sigma&_{X\ov{X}\to f\ov{f}} = 
\frac{4\pi Q'^2_X Q'^2_f  \alpha'^2}{E^2} \frac{\beta'}{\beta} \frac{(2E^2+m_f^2)(2E^2+m_X^2)}{(4E^2-m_{Z'}^2)^2+m_{Z'}^2\Gamma_{Z'}^2} .
\end{align}
In the non-relativistic limit, the s-wave contribution dominates. As for scalar CDM, we assume Higgs exchange contributions are suppressed either because the Higgs masses are large or because the $X$-Higgs couplings are small.

The $Z'$ width $\Gamma_{Z'}$ is given by Eq.~(\ref{eq:width}) with the $ m_{Z'}^2 -4m_X^2 $ in the $X$ contribution replaced by $ m_{Z'}^2 + 2m_X^2 $.

As for the scalar CDM case, for $m_X>m_{Z'}$ there is an extra contribution from annihilations into pairs of $Z'$ bosons with cross section
\begin{align}
&\sigma_{X\ov{X}\to Z'Z'} = \frac{\pi Q'^4_X \alpha'^2}{E^2} \frac{w}{v}
\left[ -1 -\frac{(2+z^2)^2}{(2+2v^2-z^2)^2-4vw}
\right. \nonumber \\ & \left.
-\frac{6-2z^2+z^4+12v^2+4v^4}{2vw(1+v^2+w^2)} 
\ln\frac{1+(v+w)^2}{1+(v-w)^2}
\right].
\end{align}

\section{Direct detection}
\label{sec;direct}
Since we assume that the U(1)$'$ gauge boson couples to the SM 
fermions vectorially, the direct detection rate has a
spin-independent component. Since the interaction is described by vector-current operators in the microscopic theory, the effective coupling of the $Z'$ with protons, neutrons and nuclei can be obtained using the conservation of the U(1)$'$ charge. For a nucleus $N$ of mass number $A$ and electric charge $Z$, one has the effective lagrangian term
\beq
Q'_NZ'_{\mu} \ov{N} \gamma^{\mu} N,
\eeq
where
\begin{align}
Q'_N=Z Q'_p + (A-Z) Q'_n
\end{align}
is the U(1)$'$ charge of the nucleus, and
\begin{align}
 \quad Q'_n=Q'_u +2 Q'_d, \quad Q'_p=2Q'_u+Q'_d
\end{align}
are the U(1)$'$ charges of the neutron and proton, respectively. For our choice of $Q'_f=1/3$, we have $Q'_N=A$.
The non-relativistic limit of the spin-independent cross section for direct detection then follows as
\begin{equation}
\label{direct}
\sigma_{XN} = \frac{16 \pi \alpha'^2}{m_{Z'}^4}~Q'^2_X Q'^2_N
\left(  \frac{m_X m_N}{m_X + m_N}  \right)^2,
\end{equation}
where  $m_N$ is the mass of the nucleus. 

Eq.~(\ref{direct}) directly constrains the value of $m_{Z'}/g'$ once $m_X$ and $\sigma_{Xp}$ are determined in direct dark matter detection experiments. For example, for $Q'_N=1$ and $Q'_X\sim 1$, 
the DAMA/CoGeNT region around $m_X \sim 7$ GeV
and $\sigma_{Xp} \sim 10^{-40}$ cm$^2$ leads to $m_{Z'}/g'\sim1$ TeV.   

For simplicity, and as benchmark for our discussion, we take the DAMA/CoGeNT region outlined in \cite{Hooper:2010uy}, shown in orange on the $m_X$--$\sigma_{Xp}$ plane in Fig.~\ref{fig1}. Fig.~1(a) corresponds to a Dirac fermion $X$, Fig.~1(b) to a scalar $X$ (the direct detection constraints are identical in the two panels). Other analyses of the CoGeNT exponential excess (e.g.~\cite{Aalseth:2010vx,Savage:2010tg,Schwetz:2010gv}) recover different regions in the $m_X$--$\sigma_{Xp}$ plane, mostly to the right of the orange region shown, some compatible and some incompatible with the DAMA/LIBRA modulation region.

Fig.~\ref{fig1} also shows the best current bounds from negative dark matter searches (the excluded region in blue): CRESST at lower masses (from \cite{Savage:2010tg}) and XENON10 at larger masses. Two curves are shown for the XENON10 bound, one from \cite{Aprile:2011}, the other from \cite{Savage:2010tg}. They reflect different assumptions on the light detection efficiency near the threshold of the detector: the assumption with the higher detector sensitivity excludes the DAMA/CoGeNT region, the other does not.

\section{Relic density}
\label{sec;relic}

The thermal density of the CDM particles $X$ is given by the Boltzmann equation,
\beq
\frac{d n}{dt} +3Hn =- \langle \sigma_{\rm ann} v \rangle (n^2-n^2_{eq} ), 
\eeq
where $n$ is the $X$ number density and $n_{eq}$ is its value in thermal equilibrium. 

To compute the relic density, we use the procedure in \cite{Gondolo:1990dk} as implemented in DarkSUSY \cite{Gondolo:2004sc}. For this purpose, we introduced into DarkSUSY the invariant annihilation rate $W=8Ep\sigma_{\rm ann}$, where $\sigma_{\rm ann} = \sum_f \sigma_{XX^\dagger\to ff} +\sigma_{XX^\dagger\to Z'Z'} $ is the total $XX^{*}$ or $X\ov{X}$ annihilation cross section given above.

We impose that the computed cosmic density of $X$ particles $\Omega_X h^2$ (in units of $1.8783\times10^{-26}$ kg/m$^3$) equals the observed value of the cold dark matter density $\Omega_c h^2=0.1123\pm0.0035$ \cite{Komatsu:2010fb}. 
The thermal relic density depends on $\alpha'$, $m_{Z'}$ and $m_X$. 
If we fit the DAMA/CoGeNT region, the resulting parameters 
$\alpha'$ and $m_{Z'}$ lead to a thermal density that is too large 
unless the annihilation is close, but not too close, to the resonance at $m_{X} \approx m_{Z'}/2$.

\begin{figure}
\includegraphics[width=0.45\textwidth]{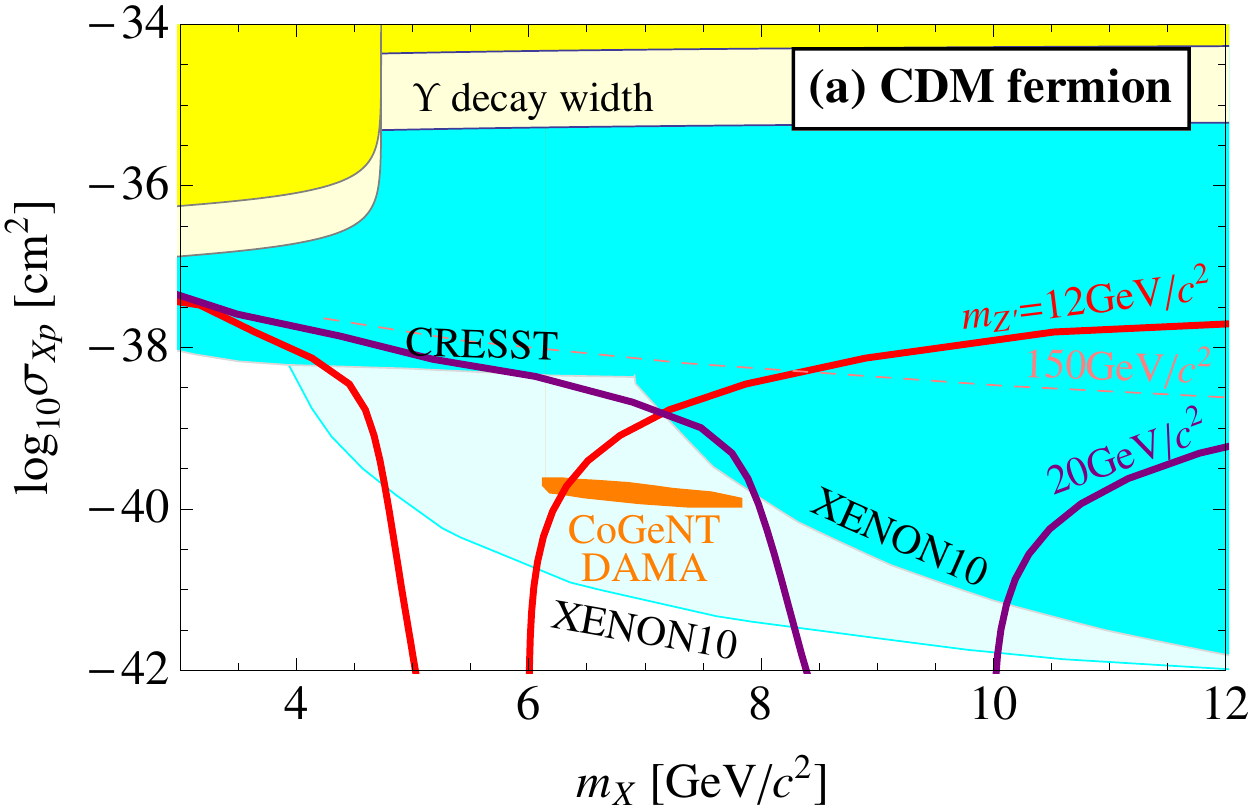}
\includegraphics[width=0.45\textwidth]{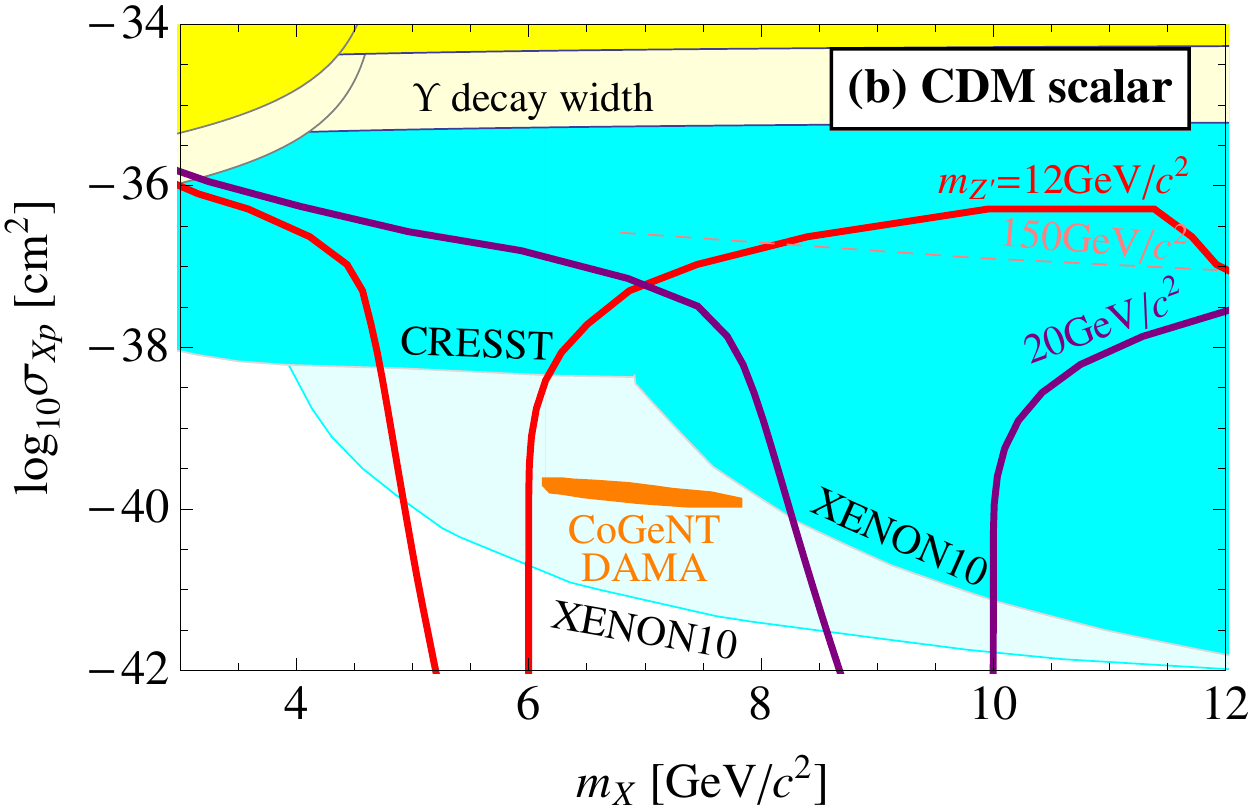}
\caption{Contour lines of $\Omega_X=\Omega_c$ for several values of the $Z'$ boson mass $m_{Z'}$. On each contour, the cosmic density of particles $X$ (fermions in panel (a), and scalars in panel (b)) equals the cosmic density of cold dark matter. Also shown are the DAMA/CoGeNT region (in orange), direct detection constraints (in blue), and accelerator constraints (in yellow).}
\label{fig1}
\end{figure}

Contour lines of $\Omega_X=\Omega_c$ in the $m_X$--$\sigma_{Xp}$ plane are shown in Fig.~\ref{fig1} for several values of $m_{Z'}$ (the error bars on $\Omega_c h^2$ are within the thickness of the lines drawn). The parameter $\alpha'$ changes along each line. Below each line, one has $\Omega_X>\Omega_c$. 
The thick red and purple contours correspond to $m_{Z'}=$12 GeV$/c^2$ and 20 GeV$/c^2$, respectively. Each contour shows a dip at $m_X=m_{Z'}/2$ due to the annihilation through the $Z'$ resonance. As a function of $m_X$, the resonance dip is highly asymmetric, being wider at $m_X<m_{Z'}/2$. This is the correct behavior expected from the finite-temperature momentum distribution of particles $X$ during annihilation in the early universe \cite{Gondolo:1990dk}.

\begin{figure}
\includegraphics[width=0.45\textwidth]{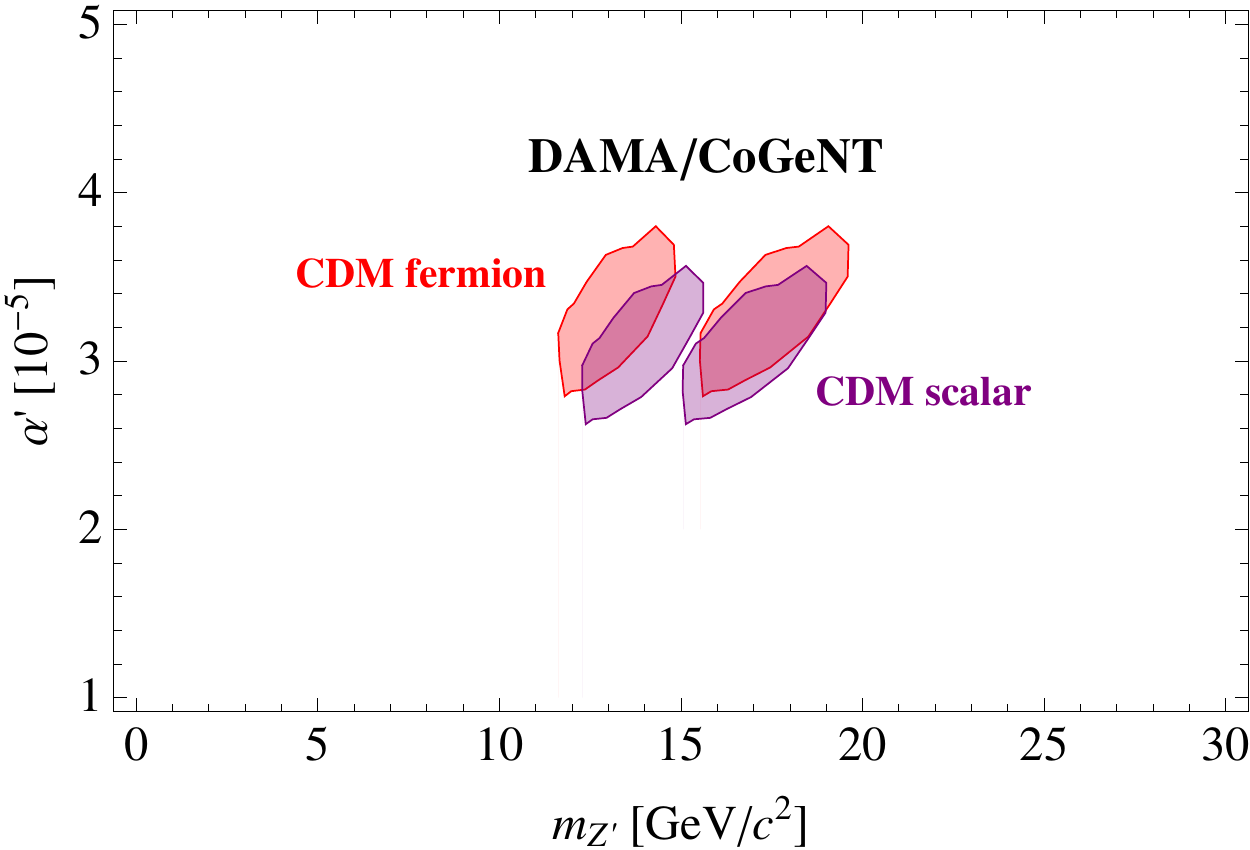}
\caption{U(1)$'$ gauge boson mass $m_{Z'}$ and coupling constant $\alpha'$ that can explain the DAMA/CoGeNT region with a light cold dark matter scalar (in purple) or Dirac fermion (in red). Regions on the right have $m_X$ on the ``left'' of the resonance ($m_X<m_{Z'}/2$); regions on the left have $m_X>m_{Z'}/2$.}
\label{fig2}
\end{figure}

We see that the $\Omega_X=\Omega_c$ contour lines sweep the DAMA/CoGeNT region for $Z'$ masses in the range $\sim1$ to $\sim20$ GeV$/c^2$, touching the DAMA/COGeNT region on the left at the lowest $m_{Z'}$ and on the right at highest $m_{Z'}$. Fig.~\ref{fig2} gives a better visualization of the range of masses $m_{Z'}$ and coupling constants $\alpha'$ that fit the DAMA/CoGeNT region.

Notice that a heavy $Z'$ with $m_{Z'}\sim$150 GeV$/c^2$, such as in suggested explanations of the CDF $Wjj$ anomaly, has trouble matching the DAMA/CoGeNT region. If such a heavy $Z'$ couples universally to quarks, $\Omega_X$ in the DAMA/CoGeNT region would be too high, as seen by the location of the 150-GeV$/c^2$ dashed line in Fig.~\ref{fig1}. A correct $X$ density may be obtained with non-universal couplings to quarks $Q'_b \gg Q'_u,Q'_d$, as summarily assumed in \cite{Hooper:2010uy}, but such non-universal couplings may be very difficult to implement in a viable model without violating constraints from, for example, flavor changing neutral currents.

\section{Accelerator bounds}

As discussed above, direct detection requires $m_{Z'}/g'\sim1$ TeV$/c^2$, while the $\Omega_X=\Omega_c$ constraint leads to $m_X\sim$10--20 GeV$/c^2$. It follows that $g' \sim 10^{-2}$, which is small but not unreasonably small. This region of small $g'$ and small $m_{Z'}$ is hard to reach in accelerator experiments, specifically because by assuming a leptophobic $Z'$ we have avoided otherwise strong experimental constraints from LEP-II and the Tevatron. For U(1)'=U(1)$_B$, the strongest bounds come from the invisible and hadronic decay widths of the $\Upsilon$ meson \cite{Carone:1995,Aranda:1998fr,Fayet:2009tv}. The region excluded by these bounds is shown in yellow in Fig.~\ref{fig1}, the edges of the yellow region corresponding to the $Z'$ masses plotted in the figure, namely $m_{Z'}=12$ GeV/$c^2$ (lower edge) and $m_{Z'}=20$ GeV/$c^2$ (upper edge). The constraint from the invisible $\Upsilon$ width \cite{Fayet:2009tv} is stronger at small $m_X$, that from the hadronic width \cite{Aranda:1998fr} at larger $m_X$.  Clearly, accelerator bounds have no effect on our scenarios.

\

%



\section{Summary}
\label{summary}

We have presented a proper quantitative analysis of two viable models for a light dark matter particle that can account for the CoGeNT and DAMA/LIBRA experimental results, i.e.\ have mass $m_X\sim7$ GeV$/c^2$, scattering cross section with nucleons $\sigma_{Xp} \sim 10^{-40}$ cm$^2$, and cosmic density equal to the cosmic density of cold dark matter. In one model the dark matter particle is a scalar, in the other a Dirac fermion. Both models assume that the interaction of the dark matter particles with ordinary matter occur through the exchange of a new leptophobic gauge boson $Z'$, and that the dark matter is produced thermally in the early universe. 

We find  viable scenarios in which the $Z'$ boson is light, with mass $m_{Z'} \sim 10$--20 GeV$/c^2$, and gauge coupling constant $g'\sim0.02$, or $\alpha'\sim10^{-5}$, which is smaller than the Standard Model gauge coupling constants but not unreasonably small. The small values of $m_{Z'}$ and $g'$ make accelerator constraints on our leptophobic $Z'$ models ineffective.

\begin{acknowledgments}
The work of PK is supported in part by SRC program of National Research 
Foundation, Seoul National University, KNRC.  PG has been partially supported by NSF award PHY-0756962 at the University of Utah, and thanks KIAS for kind hospitality during the course of this work.
\end{acknowledgments}


\end{document}